\documentstyle[twoside,fleqn,espcrc2]{article}
\input{psfig}

\newcommand{\AmS}{{\protect\the\textfont2
  A\kern-.1667em\lower.5ex\hbox{M}\kern-.125emS}}

\title{Study of Critical Slowing-Down in  
         $SU(2)$ Landau Gauge Fixing}

\author{Attilio Cucchieri\address{Department of Physics, New York University,
             4 Washington Place, New York, NY 10003, USA}
        \addtocounter{address}{-1}
        $\!\!\!\mbox{and}$ Tereza Mendes\addressmark\thanks{Poster presented
        by T.~Mendes.}}
       
\begin{document}

\begin{abstract}
        We study the problem of critical slowing-down for gauge-fixing
        algorithms (Landau gauge) in $SU(2)$ lattice gauge theory on 
        $2$ and $4$ dimensional lattices, both numerically and analytically.
        We consider five such algorithms, and we measure four different
        observables. A detailed discussion and analysis of the tuning of
        these algorithms is also presented.
\end{abstract}

\maketitle

\def\spose#1{\hbox to 0pt{#1\hss}}
\def\ltapprox{\mathrel{\spose{\lower 3pt\hbox{$\mathchar"218$}}
 \raise 2.0pt\hbox{$\mathchar"13C$}}}
\def\gtapprox{\mathrel{\spose{\lower 3pt\hbox{$\mathchar"218$}}
 \raise 2.0pt\hbox{$\mathchar"13E$}}}
\def\inapprox{\mathrel{\spose{\lower 3pt\hbox{$\mathchar"218$}}
 \raise 2.0pt\hbox{$\mathchar"232$}}}

\newcommand{\1}{1\!\!\!\bot}
\def\bgamma{\vec{\mbox{$\gamma$}}}
\def\bt{\vec{\mbox{$t$}}}
\def\bvu{\vec{\mbox{$u$}}}
\def\bvg{\vec{\mbox{$g$}}}
\def\bvv{\vec{\mbox{$v$}}}
\def\bsig{\vec{\mbox{$\sigma$}}}
\def\bx{\mbox{\protect\boldmath $x$}}
\def\by{\mbox{\protect\boldmath $y$}}
\def\bun{\mbox{\protect\boldmath $e$}}
\def\bu{\mbox{\protect\boldmath $u$}}
\def\bk{\mbox{\protect\boldmath $k$}}
\def\bp{\mbox{\protect\boldmath $p$}}
\def\bfv{\vec{\mbox{$f$}}}

\font\german=eufm10 scaled\magstep1     
\def\germang{\hbox{\german g}}
\def\germansu{\hbox{\german su}}
\def\germanso{\hbox{\german so}}

\newcommand{\be}{\begin{equation}}
\newcommand{\ee}{\end{equation}}

\def\bsigma{\mbox{\protect\boldmath $\sigma$}}
\def\btau{\mbox{\protect\boldmath $\tau$}}
\def\brho{\mbox{\protect\boldmath $\rho$}}
\def\ba{\mbox{\protect\boldmath $a$}}
\def\br{{\bf r}}

\newcommand{\tr}{\mathop{\rm tr}\nolimits}

\newcommand{\reff}[1]{(\ref{#1})}

 
 
\font\srm=cmr7                             
\font\tenmsx=msxm10 scaled\magstep1        
\font\specialroman=msym10 scaled\magstep1  
\font\sevenspecialroman=msym7              

 
\def\Z{{\hbox{\specialroman Z}}}
\def\zed{{\hbox{\specialroman Z}}}
\def\szed{{\hbox{\sevenspecialroman Z}}}
\def\R{{\hbox{\specialroman R}}}
\def\sR{{\hbox{\sevenspecialroman R}}}
\def\C{{\hbox{\specialroman C}}}
\def\sC{{\hbox{\sevenspecialroman C}}}
\def\Cbar{{\bar{\C}}}
\renewcommand{\emptyset}{{\hbox{\specialroman ?}}}
\def\restrict{{\hbox{\tenmsx \char"016}}}
 

\section{INTRODUCTION}

Here we report on the status of our study \cite{CM,CM2} of
Landau gauge-fixing algorithms.
The efficiency of these algorithms is of key importance
when gauge-dependent quantities --- such as gluon and quark propagators
--- are evaluated, especially if the effect of Gribov copies
is considered.

The main issue regarding the efficiency of these algorithms is the
problem of {\em critical slowing-down} (CSD), which
occurs when 
the {\em relaxation time} $\tau$
of an algorithm diverges as the lattice
volume is increased (see for example \cite{CSD}). Conventional local 
algorithms have {\em dynamic critical exponent} $z\approx 2$,
namely $\tau \sim N^2$, while
global methods may succeed in eliminating CSD completely,
{\em i.e.}\ $z\approx 0$.
We consider
five different algorithms:
the {\em Los Alamos} method (a conventional local
algorithm), the {\em Cornell} method, which is 
generally believed to have $z\approx 2$, the {\em overrelaxation} and
{\em stochastic overrelaxation} methods (improved local algorithms,
which are expected to show $z\approx 1$), and the 
{\em Fourier acceleration} method, which is a global method.
We confirm the predictions for $z$ (in both two and four dimensions)
with the exception of the Cornell method.
For this case we obtain that the dynamic critical exponent
is actually $z\ltapprox 1$, a result
that can be understood by a comparative
analysis between the Cornell and the overrelaxation methods.
Besides the problem of CSD, we are also
interested in understanding which quantities
should be used to test the convergence of the
gauge fixing, and in finding prescriptions for the
{\em tuning} of parameters, when needed.

We consider the Standard Wilson action in $d$ dimensions:
\be
S\left(\left\{U\right\}\right) \equiv
   \frac{4 a^{d-4}}{g^{2}_{0}} \, \frac{1}{2} \sum_{\Box}
       (1 \, - \frac{1}{2} \tr U_{\Box})
\;\mbox{.}
\ee
In order to fix the Landau gauge, we look for a {\em local}
minimum of the function \cite{W2}
\be
{\cal E}\left(\left\{ g \right\}\right) \;=\; 1 -
\frac{a^{d}}{2\,d\,V} \sum_{\bx, \, \mu} \, \tr U^{(g)}_{\mu}(\bx)
\label{eq:Emin}
\ee
where
\be
U_{\mu}^{\left( g \right)}(\bx) \; \equiv \; g(\bx) \;
          U_{\mu}(\bx) \;
                   g^{\dagger}(\bx + a \bun_{\mu})
\;\mbox{,}
\label{eq:Ug}
\ee
and we start from $g(\bx) = \1$ for all $x$,
keeping the thermalized configuration $\{ U_{\mu}(\bx) \}$ fixed.
[For the $SU(2)$ matrices we
employ the usual parametrization $U \equiv u_{0} \, \1 + i\, \bvu \cdot \bsig$,
where the components of $\bsig$ are the three
Pauli matrices.]

If $\{ U_{\mu}^{\left( g \right)}\left(x\right) \}$ is a stationary point of
${\cal E}(\{ g \})$ then we have the well-known result
\be
\left[\left(\nabla \cdot A^{\left( g \right)} \right)(x)\right]_{j} = 0 \qquad \qquad
  \forall \; \; \; x \mbox{,} \; \; j
\label{eq:diverg0}
\;\mbox{,}
\ee
namely the lattice divergence of each color
component of the gauge field
\be
A_{\mu}(\bx) \,\equiv\, \frac{1}{2 a g_{0}}
              \left[ \; U_{\mu}(\bx) -
                    U_{\mu}^{\dagger}(\bx)
              \; \right]
\ee
is null.
From equation \reff{eq:diverg0}
it follows that 
\be
Q_{\nu}(x_{\nu})  \equiv  \sum\protect_{\mu \neq \nu}
\sum\protect_{x_{\mu}}  A_{\nu}(\bx)
\ee
is constant, namely it is
independent of $x_{\nu}$.

The algorithms we consider are all (with the exception of the Fourier
acceleration) based on updating a single-site variable $g(\by)$ at a
time. As a function of $g(\by)$ only, the minimizing function becomes
\be
{\widetilde {\cal E}}\left[ g(\by) \right] \, = \, constant \,-\,
 \frac{a^{d}}{2\,d\,V} \, \tr [g(\by) \, h(\by)]
\;,
\ee
where 
\begin{eqnarray}
h(\by) &\equiv&
        \sum_{\mu}
            \left[ \; U_{\mu}(\by) \;
                 g^{\dagger}(\by + a \bun_{\mu}) \right. \nonumber \\
&  & \quad  +\, \left.   U_{\mu}^{\dagger}(\by - a \bun_{\mu}) \;
        g^{\dagger}(\by - a \bun_{\mu}) \; \right] 
\end{eqnarray}
is the {\em single-site effective magnetic field}.
Note that $h(\by)$ is proportional to an $SU(2)$ matrix, {\em i.e.}\ we
can write
\be
h(\by) \, \equiv \, {\cal N}(\by) \, {\widetilde h}(\by) \, \equiv
 \,\sqrt{\det h(\by)} \, {\widetilde h}(\by)
\;\mbox{.}
\ee
We also define 
${\cal T}(\by) \, \equiv\,  \tr [\, g(\by) \, {\widetilde h}(\by)\, ]$.

The single-site (multiplicative) update can be written
as
\be
g^{(new)}(\by)
\equiv R^{(update)}(\by) \, g^{(old)}(\by)
\;\mbox{,} \nonumber
\ee
and for the methods we consider we have (see \cite{CM} and references
therein):

\noindent {\bf Los Alamos Method:} in this case we
have
$g^{(new)}(\by) \,=\, {\widetilde h}^{\dagger}(\by)$,
{\em i.e.}\ this update brings the single-site
function ${\widetilde {\cal E}}\left[ g(\by)
\right]$ to its unique absolute minimum;

\noindent {\bf Overrelaxation:} here the matrix
$R^{(update)}(\by)$ is given by  $[{\widetilde h}^{\dagger}(\by)\,
g^{\dagger}(\by)]^{\omega} $,
with $\omega\in (1,2)$; notice that the Los Alamos method
corresponds to the case in which $\omega$ is equal to one, and
that for $\omega = 2$ the value of
${\widetilde {\cal E}}\left[ g(\by)
\right]$ does not change;

\noindent {\bf Stochastic Overrelaxation:} in this case
the update $g^{(new)}(\by)$ is given by
$$
\left[{\widetilde h}^{\dagger}(\by) \,g^{\dagger}(\by)
\, \right]^{2}\,  g(\by)\, = \,{\widetilde h}(\by)\, {\cal T}(\by)
\,-\, g(\by) \qquad \qquad 
$$
with probability $p$, and
by ${\widetilde h}^{\dagger}(\by)$ with probability $1 - p\, $;

\noindent {\bf Cornell Method:} here
$R^{(update)}(\by)$ is proportional to $\left[ \1 - \alpha \,
 a^{2}\, g_{0}\, \left( \nabla \cdot A^{\left( g \right)} \right) (\by)
\right]$; in \cite{CM} we prove that this can also be
written as
$$
\left[\, 1 \, - \, \frac{ \alpha \,{\cal N}(\by)\, {\cal T}(\by) }{ 2}
   \,\right]
  \, \1 \, + \, \alpha \, {\cal N}(\by)\, {\widetilde h}^{\dagger}(\by)\, g^{\dagger}(\by)
\;\mbox{;} \,\,\,\,\,\,
$$

\noindent {\bf Fourier Acceleration:} in this case
the matrix
$R^{(update)}(\by)$ is proportional to
$$
\1\, -\, \left\{ {\widehat F}^{-1}
         \, \frac{ p^{2}_{max} }{ p^{2}(\bk) } \,
                {\widehat F} \left[ \, \alpha \,
 a^{2}\, g_{0}\, \left( \nabla \cdot A^{\left( g \right)} \right)
\right]\, \right\}(\by) \;\mbox{,}
$$
where ${\widehat F}$ is the Fourier transform,
$p^{2}$ is defined as $ (4/a^2) \, \sum_{\mu} \,
\sin^{2}\left( \, \pi \,a\,k_{\mu} \, \right)$,
and $\,a\,k_{\mu} N$ takes the values $0\mbox{,}\,1\mbox{,}\,\ldots
\mbox{,}\,N-1$. (Of course, to reduce the number of
times the Fourier transform is evaluated, a checkerboard
update should be employed.)

In order to determine the convergence of the gauge fixing, we measure
the relaxation times $\tau_i$ for the following quantities \cite{CM}:
\begin{eqnarray}
{e_{1}}(t) \!\! &\equiv& \!\! {\cal E}(t - 1) - {\cal E}(t) \nonumber
\\[1mm]
{e_{2}}(t)\!\!  &\equiv&\!\! \frac{a^{d+4}\, g^{2}_{0}}{V} \sum_{\bx,\,j} \,
 \Big[ \left( \nabla \cdot A \right) (\bx) \Big]_{j}^{2} \nonumber
\\
{e_{4}}(t)\!\!  &\equiv&\!\! \max_{\bx} \left[ \, 1 \, - \, \frac{1}{2} \,
   \tr\,R^{(update)}(\bx) \right] \nonumber
\\
{e_{6}}(t)\!\!  &\equiv& \!\! \frac{1}{3d\,N} \,
  \sum_{\nu,\,j,\,x_{\nu}} \,
    \left[ \,  Q_{\nu}(x_{\nu}) - {\widehat Q}_{\nu}  \,
      \right]_{j}^{2} \left[ {\widehat Q}_{\nu} \right]_{j}^{-2} \nonumber
\end{eqnarray}
Here $t$ indicates the number of sweeps of the lattice and
$ {\widehat Q}_{\nu} \,
\equiv \, 1/N \, \sum_{x_{\nu} = 1}^{N} \,
            Q_{\nu}(x_{\nu}) $.
We expect to observe that
\be
e_{i}(t) \, \propto \,\exp\left(\, - \, t / \tau_{i} \, \right)
\quad \mbox{with} \quad
\tau_i = c_i N^{z_i}
\;\mbox{,}
\ee
and also that all the quantities above have the same relaxation time $\tau$,
and hence the same $c$, $z$.

\section{RESULTS FOR $d=2$}

For the four local methods we used \cite{CM}
lattice sizes $N = 8\mbox{,}\, 12\mbox{,}\,\ldots
\mbox{,}\,36$, while in the Fourier acceleration case
we considered $N = 8\mbox{,}\, 16\mbox{,}\,32\mbox{,}\,64$. In all
cases we have chosen the {\em constant physics}
$N^{2} / \beta = 32$, namely $N / \xi \approx 7$.
We stopped the gauge fixing when the condition
$e_{2} \leq 10^{- 12}$ was satisfied.

From our data it is clear that
the four quantities $e_{i}$ have the same $\tau$
for each given algorithm, as expected.
However, for all the local updates
--- with the exception
of the stochastic overrelaxation ---
the quantity $e_6$ was hardest to relax, namely its
value was several orders of magnitude larger
than the value of the other quantities.
On the contrary, the Fourier acceleration method
seems to be very efficient in relaxing $e_6$.
Basing on these results,
we think that the quantity $e_6$ allows a very
sensible check of the gauge fixing and, in our opinion,
it should always be used.

We evaluate the dynamic critical exponents $z$ from the 
weighted least-squares fit for $\tau = c \,
N^{z}$, using lattice sizes $N \geq N_{min}\;$,
and obtain \cite{CM}:
\begin{eqnarray}
\mbox{{\bf Los Alamos}} & z = 1.99\pm 0.04 \;\mbox{,} & \! N_{min} = 12 \nonumber \\
\mbox{{\bf Cornell}} & z = 0.83\pm 0.09 \;\mbox{,} & \! N_{min} = 16 \nonumber \\
\mbox{{\bf Overrelax.}} & z = 1.12 \pm 0.07\;\mbox{,} & \! N_{min} = 16 \nonumber \\
\mbox{{\bf Stoch. Overr.}} & z = 1.09\pm 0.05 \;\mbox{,} & \! N_{min} = 12 \nonumber \\
\mbox{{\bf Fourier}} & z = 0.04 \pm 0.06\;\mbox{,} & \! N_{min} = 8 \nonumber
\end{eqnarray}

Clearly, the {Fourier acceleration} method is the most
successful in reducing CSD, and 
it should be the method of choice when large lattice sizes are employed.
However, the method is more costly per iteration than the {local methods},
and its CPU-time/site increases logarithmically with the lattice
size. When this is taken into account, the total time for gauge fixing
a configuration is smaller for Fourier acceleration
than for the local methods if the lattice size is larger
than about 300 sites, as can be seen in Figure 1. [Of course this
result is very machine- and code-dependent. In any case, it seems unlikely
\cite{CM} that
the Fourier acceleration method would become the method
of choice in two dimensions at lattice sizes $N$ smaller than around 100 sites.]

\begin{figure}
\psfig{figure=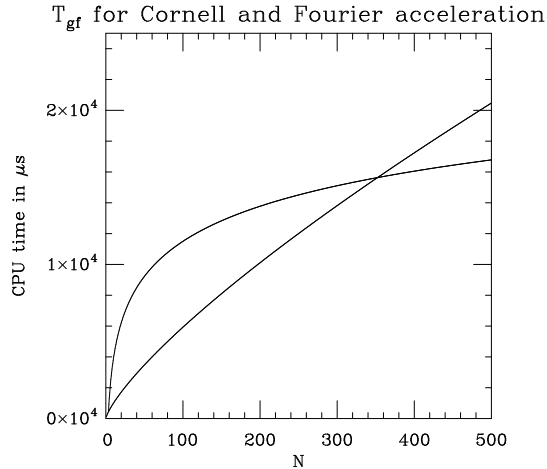,height=2.5in}
\vskip -0.8cm
\caption{Comparison of the ``effective'' CPU-time/site
($T_{gf}$) between the Cornell method 
(the nearly straight line) and the Fourier acceleration.}
\end{figure}

\section{TUNING}

The need for tuning the parameters of an algorithm in order to get
optimal efficiency is, of course, a potential disadvantage. Of the five
algorithms we consider, all (but the Los Alamos method) require
tuning. We did a careful analysis of this problem and
we were able to verify a simple analytic expression
for the optimal choice of $\omega$
(overrelaxation method), and to relate $\omega_{opt}$ to the optimal choice
for the parameters of the other {local methods}. More specifically:

\noindent {\bf {Overrelaxation}:} We tested the Ansatz \cite{MO3}
\be
\omega_{opt}\, = \, 2\, \left[\, 1 \, + \, C_{opt} / N\, \right]^{- 1}
\ee
and obtained very good agreement for lattice sizes greater than 12
with $C_{opt} = 1.53 \pm 0.35$. Note that,
as the number of iterations $t$ increases, the matrix $R^{(update)}$
should approach the identity matrix $\1$. In this limit,
the overrelaxation update can be written as \cite{CM}
\be
g^{(over)}(\by) \, \propto \,
( 1 - \omega ) \, g(\by)\, +\, \omega
      \, {\widetilde h}^{\dagger}(\by) \;.
\label{eq:gover}
\ee

\noindent {\bf {Cornell Method}:} In this case, in the limit
of large $t$, we can write \cite{CM}
$$
g^{(Corn)}(\by) \, \propto \,
[ 1 - \alpha\, {\cal N}(\by) ] \, g(\by)\, +\, \alpha
        \,{\cal N}(\by)
      \, {\widetilde h}^{\dagger}(\by) \;\mbox{.}
$$
By comparing this equation with \reff{eq:gover},
we made the conjecture
{$\alpha_{opt}\,\langle\,{\cal N}\,\rangle\,=\,\omega_{opt}$},
with $\langle\,{\cal N}\,\rangle$ given by
$2d \,(1\,-\,\langle\,{\cal E}_{min}\,\rangle) \;$.
This Ansatz is very well fitted by our data \cite{CM}.

\noindent {\bf {Stochastic Overrelaxation:}} In this case
it is not clear how to write a formula for the update as
the number of iterations $t$ increases. A possibility
\cite{CM} is to consider
\be
g^{(stoc)}(\by) \, \approx \, \left(\,1 + p\right) \,
         {\widetilde h}^{\dagger}(\by) \, - \, p \, g(\by)
\;\mbox{,}
\ee
which suggests the relation
{$p_{opt}\,=\,\omega_{opt}\,-\,1$}.
This conjecture seems to be
satisfied reasonably well for lattice sizes larger than 30.

\noindent {\bf {Fourier Acceleration:}} By analogy with
the continuum $U(1)$ case, we expect \cite{DBKKLWRS}
\be
a^{2}\,g_{0}\,(\nabla\cdot A^{\left( g \right)})(\by)
\propto
\exp \left[  - \alpha\,t \,a^{2}
p^{2}_{max} \right] \; \mbox{,}
\ee 
and therefore with the choice $\alpha_{opt} = ( a \, p_{max} )^{- 2}\,$
we should obtain $\tau \propto{\cal O}(1)$. This gives,
in $d = 2$, a value ${\alpha_{opt} \,=\,0.125}$.
This result is only in qualitative agreement with our data, which seem
to indicate that $\alpha_{opt} \approx 0.16$
for large lattice sizes.

\section{EXTENSION TO $d=4$}

We study (in two and in four
dimensions) the case $\beta = \infty$, namely
we fix $U_{\mu}(\bx) = \1$ in \reff{eq:Ug} for all $\bx$ and $\mu$,
and we look for a local minimum of the function \reff{eq:Emin}
starting from a random configuration $\{ g(\bx) \}$.
In both the two and four dimensional cases,
we essentially confirm the results obtained for $z$ at finite
$\beta$ and in two dimensions \cite{CM2}.

It is interesting to notice that, in the case
$\beta = \infty$, we can also write the minimizing function
as
\be
{\cal E}\left(\left\{ g \right\}\right) \,=\,
\frac{a^{d}}{2 \, d \, V} \sum_{\bx, \, \mu}
 \| \, g(\bx)\,-\,g(\bx + a \bun_{\mu})\,\|^{2}
\mbox{.}
\ee
Here the $SU(2)$ matrix $g = g_{0} \1 + i\, \bsig \cdot \bvg$
is considered as a four-dimensional unit vector $( g_{0}\mbox{,}\,
\bvg )$
and $\| g \|^{2} \equiv g_{0}^{2} + \bvg \cdot \bvg$.
When we are close to a minimum we can write
\be
g(\bx) \, = \, \1 -\, i\, \epsilon \,\bsig\cdot\bfv(\bx)
         \,+\,{\cal O}(\epsilon^2)
\;\mbox{,}
\ee
where $\bfv(\bx)$ is a three-vector field.
This gives
\be
g(\bx)\,g^{\dagger}(\bx)\,=\,\1\,+\,{\cal O}(\epsilon^2)
\ee
and
\be
{\cal E}\left(\left\{ g \right\}\right)
\, \approx \, \frac{\epsilon^{2} \, a^{d}}{2\, d\, V}\sum_{\bx, \, \mu}
\| \, \bfv(\bx)\,-\,\bfv(\bx + a \bun_{\mu})\,\|^{2}
\mbox{,}
\ee
namely (up to order $\epsilon^{2}$) we have the action of a three-vector
massless free field $\bfv(\bx)$. Therefore we can
use standard analytic
methods \cite{N} in order to study the problem of
minimizing this quadratic form, and
to compare the results for $z$, $\tau$ and the optimal choice of the
parameters with our numerical data.
Indeed, we find good agreement for all the quantities and methods
considered \cite{CM2}.

Finally, we extend our simulations to the {case
$\beta = 0$} (again in two and in four
dimensions) \cite{CM2}.
All our results at $\beta > 0$ are confirmed
except for the Cornell method, for which we get $z\approx 2$,
and for the Fourier method, which shows $z\approx 1$.
In the Cornell case this
result can be easily understood,
since at $\beta = 0$
its single-site update does not decrease the minimizing function
at each step. In particular, the value of ${\cal E}$ {\em increases}
if $\alpha {\cal N}(\by)$ is large than two. [This is obvious if we consider
the analogy between the Cornell method and the overrelaxation method,
and the relation $\alpha {\cal N}(\by) \sim \omega$. Moreover, it
is plausible that only at very small values of $\beta$ we can have
$\alpha \langle {\cal N} \rangle$ smaller than two and, at the same
time, $\alpha {\cal N}(\by) > 2$ for a large set of lattice sites $\by$.]
However, we can recover
the value $z\approx 1$ by redefining the Cornell single-site update in the
following way \cite{CM2}
$$
g^{(Corn)}(\by)  \propto 
\left[ 1 - \frac{ \tilde{\alpha}(\by) {\cal T}(\by) }{ 2 } \right]
 g(\by)\, +\, \tilde{\alpha}(\by)
      \, {\widetilde h}^{\dagger}(\by) \,\mbox{,}
$$
with $ \tilde{\alpha}(\by) \equiv \min\left(\,\alpha\, {\cal
N}(\by)\mbox{,}\;2\,\right) $.
In the Fourier acceleration case we have the same problem:
the update can increase the value of the minimizing
function ${\cal E}$, and this happens most likely for
small values of $\beta$. However, in this case
we did not find a redefinition of the
Fourier update \cite{CM2} which could recover the
value $z \approx 0$.


\begin{thebibliography}{9}

\bibitem{CM} A.\ Cucchieri and T.\ Mendes, Nucl.\ Phys.\
             {\bf B471} (1996) 263. 

\bibitem{CM2} A.\ Cucchieri and T.\ Mendes, {\em Critical Slowing-Down in
             Four-Dimensional
             $SU(2)$ Landau Gauge Fixing Algorithms}, in preparation.

\bibitem{CSD} L.\ Adler, Nucl.\ Phys.\  {\bf B} (Proc.\ Suppl.)
              9 (1989) 437;
              U.\ Wolff, Nucl.\ Phys.\  {\bf B} (Proc.\ Suppl.)
              17 (1990) 93;
              A.\ D.\ Sokal, Nucl.\ Phys.\  {\bf B} (Proc.\ Suppl.)
              20 (1991) 55.

\bibitem{W2} K.\ G.\ Wilson, {\em Recent Developments in Gauge Theories}
             Proc.\  NATO Advanced Study Institute (Carge\`se, 1979),
             eds.\  G.\  't Hooft et al.\  (Plenum Press, New York--London,
             1980).

\bibitem{MO3} J.\ E.\ Mandula and M.\ Ogilvie, Phys.\ Lett.\  {\bf B248} (1990)
              156.

\bibitem{DBKKLWRS} C.\ T.\ H.\ Davies et al., Phys.\ Rev.\
                   {\bf D37} (1988) 1581.

\bibitem{N} H.\ Neuberger, Phys.\ Rev.\ Lett.\  {\bf 59} (1987) 1877;
            U.\ Wolff, Phys.\ Lett.\  {\bf B288} (1992) 166.

\end{thebibliography}
\end{document}